\documentclass[aps,prl,floats,amsmath,showpacs,twocolumn]{revtex4}
\usepackage{graphicx}
\usepackage{multirow}

\begin{document}

\title{Pressure-induced structural, electronic, and magnetic effects
in BiFeO$_3$}

\author{O.E. Gonz\'alez-V\'azquez and Jorge \'I\~niguez}

\affiliation{Institut de Ci\`encia de Materials de Barcelona
(ICMAB-CSIC), Campus UAB, 08193 Bellaterra, Spain}

\begin{abstract}
We present a first-principles study of multiferroic BiFeO$_3$ at high
pressures. Our work reveals the main structural (change in Bi's
coordination and loss of ferroelectricity), electronic (spin crossover
and metallization), and magnetic (loss of order) effects favored by
compression and how they are connected. Our results are consistent
with the striking manifold transition observed experimentally by
Gavriliuk {\sl et al}. [Phys. Rev. B {\bf 77}, 155112 (2008)] and
provide an explanation for it.
\end{abstract}






\maketitle

Room-temperature multiferroic BiFeO$_3$ (BFO) is one of the most
intensively studied materials of the moment. BFO is among the most
promising multiferroics from the applications perspective, and the
latest results regarding its fundamental properties~\cite{catalanXX}
and the engineering possibilities it offers~\cite{jang08} continue to
fuel the interest in it.

Indeed, recent works suggest the correlations between the structural,
electronic, and magnetic properties of BiFeO$_3$ are not understood
yet. We have evidence for a large sensitivity of BFO's conductivity to
magnetic order~\cite{catalanXX,ravindran06}, a metal-insulator (MI)
transition driven by structural changes at high
temperatures~\cite{palai08}, metallicity at ferroelectric domain
walls~\cite{ramesh08}, and even a spin-glass phase below
150~K~\cite{singh08}. Yet, the results most revealing of the complex
interactions in BFO may be those of Gavriliuk {\sl et
al}.~\cite{gavriliuk-old,gavriliuk08}. These authors observed a
pressure-driven diffuse transition, occurring at room temperature in
the 40--50~GPa range, that involves a structural change, loss of
magnetic order, and metallization. The driving force behind such
transformations, tentatively attributed to a spin crossover of
Fe$^{3+}$, remains to be clarified.

Here we report on a first-principles study of the high-pressure
behavior of BFO in the limit of very low temperatures (nominally,
0~K). Our results reproduce the essential observations of Gavriliuk
{\sl et al}., thus revealing the mechanisms that can lead to a
manifold (structural/electronic/magnetic) transition in this material.

{\sl Methods}.-- We used the local density approximation
(LDA)~\cite{lda} to density functional theory as implemented in the
code VASP~\cite{vasp}, including the so-called LDA+U correction of
Dudarev {\sl et al}.~\cite{dudarev98} for a better treatment of iron's
3$d$ electrons. The technicalities of our calculations~\cite{fn-tech}
are standard, but our use of the LDA+U asks for a comment: In this
work we compared the energies of different electronic phases
(insulating/metallic) in which the Fe$^{3+}$ ions display different
spin states. Doing this accurately constitutes a challenge for any
{\sl ab initio} approach; in particular, while commonly used to study
such problems, the computationally-efficient LDA+U should be employed
with caution in this context. For this reason, we repeated all our
calculations for different values of the $U$ parameter in the 0--4~eV
range, and thus made sure our qualitative conclusions are
reliable. (Unless otherwise indicated, the reported results are for
$U=3$~eV.)

\begin{figure}[b!]
\includegraphics[width=0.8\columnwidth]{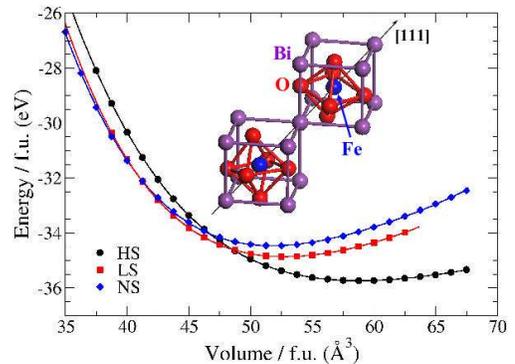}
\caption{(color online). $E(V)$ curves for the BFO phases here
considered (see text). Inset: structure of the ferroelectric $R3c$
phase of BFO stable at ambient conditions.}
\label{fig1}
\end{figure}

Phase transitions in BFO under moderately high pressures have been
studied theoretically~\cite{ravindran06}. In this work, however, we
were not interested in the relatively {\sl mild} effects so far
investigated. Rather, we wanted to determine whether pressure may
induce profound changes in the electronic structure of the
compound. Thus, we restricted our simulations to the 10-atom cell of
the $R3c$ phase of BFO stable at ambient conditions (see inset in
Fig.~\ref{fig1}), assuming this is enough to capture the phenomena of
interest. Then, as a function of volume (i.e., pressure), we performed
structural relaxations for a variety of atomic, electronic, and
magnetic configurations. That allowed us to identify a large number of
possible phases and determine their relative stability and
properties. As done in other first-principles
studies~\cite{neaton05,ederer05}, we neglected the long-period spin
cycloid in BFO.

\begin{figure}[t!]
\includegraphics[width=0.7\columnwidth]{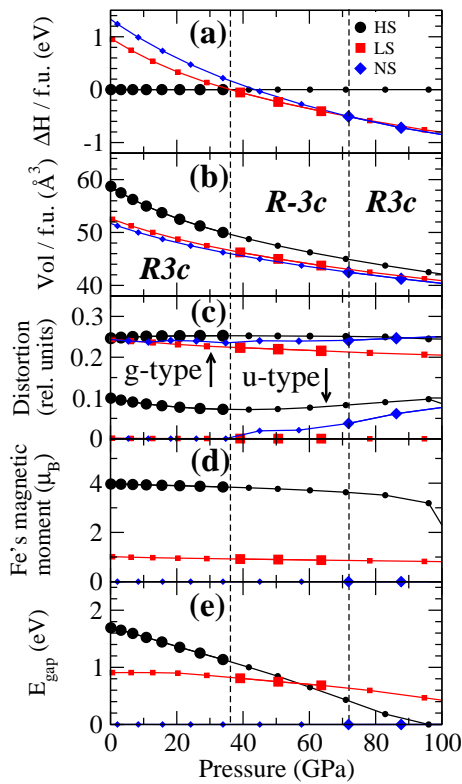}
\caption{(color online). Pressure-dependence of the enthalpy (a),
volume (b), structural distortions with respect to the ideal cubic
perovskite structure (c), Fe's localized magnetic moment (d), and
electronic band gap (e) for the phases of BFO here considered (see
text). Dashed vertical lines mark the transition pressures. Bigger
symbols indicate which phase is stable. Panel~(a): We take as the zero
of enthalpy the pressure-dependent result for the HS phase. Panel~(c):
For all phases, the upper (resp. lower) symbols correspond to the {\sl
g}-type (resp. {\sl u}-type) distortions (see text); distortions in
units relative to the pressure-dependent lattice vectors. Panel~(e):
The highlighted area is suggestive of the subtleties pertaining to the
value of $E_{\rm gap}$ in the LS phase (see text).}
\label{fig2}
\end{figure}

{\sl Structural and spin transitions}.-- Figure~\ref{fig1} displays
our results for the equation of state of the BFO phases we found to be
stable in some pressure range. The obtained pressure-driven
transitions are better visualized in Fig.~\ref{fig2}, which shows the
pressure dependence of the key properties. At ambient and moderately
high pressures we obtained the usual ferroelectric $R3c$ phase of BFO,
with G-type AFM order and Fe$^{3+}$ in a high-spin (HS) state. Then,
we found that at about 36~GPa BFO undergoes a first-order phase
transition to a phase with the Fe$^{3+}$ ions in a low-spin (LS)
configuration. The concurrent drops in volume and iron's magnetic
moment across the HS--LS transition can be seen, respectively, in
Figs.~\ref{fig2}b and
\ref{fig2}d. (The magnetic moments are nearly
pressure-independent within both the HS and LS phases.) Finally, a
metallic phase with no localized magnetic moments, denoted as ``NS
phase'' where NS stands for ``No Spin'', becomes stable above
72~GPa. Such a paramagnetic metallic phase is typical of
transition-metal oxides at high pressures~\cite{imada98}.

\begin{figure}[t!]
\includegraphics[width=0.85\columnwidth]{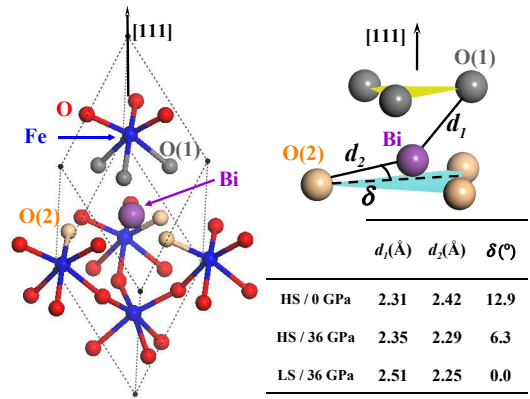}
\caption{(color online). Left: surroundings of Bi in the BFO
structure. The six nearest-neighboring oxygens are split in two
groups, denoted O(1) and O(2) and colored differently, each of which
is composed of three symmetry-related atoms that form a plane
perpendicular to the [111] direction. Right: Structural parameters
relevant to the HS--LS transition.}
\label{fig3}
\end{figure}

The main structural features of the obtained stable phases can be
described in terms of {\sl g}-type and {\sl u}-type distortions of the
ideal cubic perovskite structure, where {\sl g} (resp. {\sl u}) stands
for {\sl gerade} or {\sl even under inversion} (resp. {\sl ungerade}
or {\sl odd under inversion}). The evolution with pressure of the
distortions thus quantified is shown in Fig.~\ref{fig2}c. For all the
phases, the {\sl g}-type distortions are essentially oxygen-octahedron
rotations as those occurring in the ferroelectric $R3c$ phase of BFO
at ambient pressure; our calculations show such rotations remain
present under compression. The {\sl u}-type displacements in the HS
phase correspond to the usual ferroelectric distortion of BFO,
dominated by the stereochemical activity of bismuth: the Bi atoms move
along the [111] direction to approach the three O atoms forming a face
of the neighboring oxygen octahedron. In the LS phase, though, we find
a null {\sl u}-type distortion ($R\bar{3}c$ space group), which
reflects a change in bismuth's coordination. As shown in
Fig.~\ref{fig3}, the LS phase presents BiO$_3$ {\sl planar} groups in
which the three oxygens binding with one Bi atom belong to three
different O$_6$ octahedra. Such planar groups are also present in the
NS phase, but there they co-exist with a {\sl u}-type displacement,
along the rhombohedral axis ($R3c$ space group), in which the Bi and
Fe atoms have a quantitatively similar participation.

Thus, we found pressure favors two main modifications of the usual
$R3c$ phase, namely, a change in Bi's coordination and a low- or
null-spin configuration of the Fe$^{3+}$ ions. We also checked such
variations can exist independently (e.g., LS Fe$^{3+}$ can occur in
absence of planar BiO$_3$ groups), which gives raise to various
metastable phases. Indeed, we obtained many metastable phases (not
shown here) with different atomic and electronic structures, including
the occurrence of intermediate-spin Fe$^{3+}$. We even found a
mixed-spin phase, with one HS iron and one LS iron in our 10-atom
cell, that becomes stable in the 35--37~GPa pressure range.

{\sl Metallization at the HS--LS transition}.-- As expected, for the
HS phase we obtained an AFM insulating ground state. The energy
difference between the AFM and ferromagnetic (FM) configurations
varies from about 0.27~eV/f.u. at 0~GPa to about 0.65~eV/f.u. at
36~GPa, which is compatible with the experimentally observed high
transition temperature ($T_{\rm Neel}\approx$~643~K at 0~GPa) and its
increase with pressure~\cite{catalanXX}. As shown in Fig.~\ref{fig2}e,
the HS phase is an insulator throughout its stability range, and its
energy gap decreases with pressure.

\begin{figure}[t!]
\includegraphics[width=0.8\columnwidth]{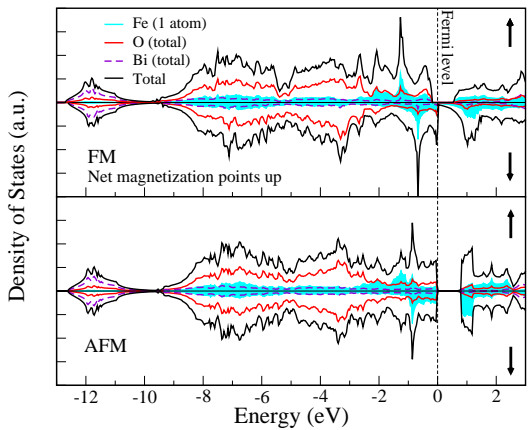}
\caption{(color online). Electronic density of states of the FM
and AFM configurations of the LS phase of BFO at 50~GPa. The shaded
area indicates the result for one Fe$^{3+}$ ion with its local
magnetic moment pointing up.}
\label{fig4}
\end{figure}

According to our results, the properties of the LS phase are more
complex. Throughout its stability range, the AFM and FM magnetic
arrangements are nearly degenerate, never differing by more than
0.03~eV/f.u. (Moreover, the ground state shifts from AFM to FM as
pressure increases.) This implies that the magnetic ordering
temperature for the LS phase will be much (about 10 times) smaller
than that of the HS phase. We also found the electronic structure of
the LS phase depends strongly on the magnetic order. Figure~\ref{fig4}
shows illustrative results at 50~GPa: The AFM case presents a gap of
about 0.8~eV, while for the FM order we get a half-metallic solution.

These results imply that, if we were able to heat the LS phase up to
room temperature ($T_{\rm room}$), we would obtain a paramagnetic
state. Further, this magnetically-disordered phase will appear as
being metallic, since the the thermally-averaged equilibrium state
will present a significant electronic density of states at the Fermi
level. Thus, our calculations suggest that, at $T_{\rm room}$, BFO
might undergo a pressure-driven HS--LS transition that brings about
simultaneous structural, magnetic, and electronic (insulator-to-metal)
transformations.

{\sl Discussion}.-- The picture of the HS--LS transition that emerges
from our simulations is essentially identical to the one suggested by
Gavriliuk {\sl et al}.~\cite{gavriliuk-old} to explain their observed
pressure-driven transition. One should keep in mind, though, that the
experiments of Ref.~\onlinecite{gavriliuk-old} were performed at
$T_{\rm room}$; thus, the identification between the experimental
transition and our HS--LS transition relies on the assumption that our
LS phase is the equilibrium state at $T_{\rm room}$.

To resolve this issue, one would need to compute the
temperature-pressure ($T$-$P$) phase diagram of BFO {\sl ab initio},
which falls beyond the scope of this work. Yet, we can estimate the
{\sl temperature} stability range of the low-$T$ phases by calculating
their enthalpy difference with the cubic phase, as such a quantity
should be roughly proportional to the temperature at which the
transition to the cubic phase occurs. At 0~GPa we obtained
0.88~eV/f.u. for the difference between the $R3c$ HS phase and the
lowest-lying cubic phase (which presents Fe$^{3+}$ in a HS state), a
large value consistent with the experimentally observed high
transition temperature (1200~K)~\cite{palai08,fn-1200}. This enthalpy
difference decreases {\sl gently} with pressure: For example, at
50~GPa, the $R\bar{3}c$ LS phase and the lowest-lying cubic phase
(which presents Fe$^{3+}$ in a LS state) differ by about
0.60~eV/f.u. Hence, our calculations suggest that, within the
stability range of the LS phase, the transition into a {\sl LS cubic
phase} will occur at temperatures well above $T_{\rm room}$ (roughly,
in the 700-800~K range). Our results thus support the hypothesis of
Gavriliuk {\sl et al}. that the metallic phase observed experimentally
at high pressures and $T_{\rm room}$ contains LS Fe$^{3+}$.

We found additional support for this identification. The computed
HS--LS transition pressure (about 36~GPa) agrees reasonably well with
the observed one (40-50~GPa), and the wealth of competing (meta)stable
phases that we found is consistent with the diffuseness of the
experimental transition. Further, the results for $V(P)$ in
Fig.~\ref{fig2}b are in good qualitative agreement with the
experimental data, and the computed bulk modulus of the HS and LS
phases are markedly different -- the HS phase being significantly
softer -- as observed experimentally for the low- and high-pressure
phases~\cite{fn-b0}. Finally, the pressure dependence of the enthalpy
difference between the low- and high-pressure phases was
experimentally determined to be 12~meV/GPa at the transition region,
and we obtain about 15~meV/GPa.

It is not our purpose here to give a detailed electronic picture of
the HS--LS transition that we found. Let us just note our results for
BFO strongly resemble what occurs in hematite (Fe$_2$O$_3$), which
undergoes a HS--LS transition with accompanying metallization at
50~GPa~\cite{brado02,rollmann04,kozhevnikov07}. Indeed, the conclusion
of Ref.~\onlinecite{kozhevnikov07} that there is an enhancement of the
metallic character of hematite's LS phase under pressure, driven by
the broadening of the not-fully-occupied $t_{2g}$ bands of Fe$^{3+}$,
seems consistent with our findings. We should also note Gavriliuk {\sl
et al}. have recently described the pressure-induced metallization in
BFO in terms of a Mott-Hubbard picture~\cite{gavriliuk08}. We have
doubts about this interpretation: First-principles calculations show
that BFO displays broad, significantly hybridized, valence bands (see
Refs.~\onlinecite{neaton05} and
\onlinecite{clark07} for the usual HS phase of BFO and Fig.~\ref{fig4}
for our LS phase), which suggests it may not be adequate to place this
material on the Mott-Hubbard side of the usual Zaanen-Sawatzky-Allen
diagram~\cite{zaanen85}.

Our results allow us to discuss currently debated features of the
$T$-$P$ phase diagram of BFO. Following the discovery of the
high-temperature metallic phase mentioned above~\cite{palai08}, which
has the ideal cubic perovskite structure with a 5-atom unit cell, it
has been proposed this phase might extend its stability range down to
low temperatures and high pressures~\cite{scott08,catalanXX}. Our
results suggest to the contrary. While restricted to a 10-atom cell,
our simulations do include the ideal cubic perovskite as a possible
solution, and show this phase does not become the ground state under
compression. For pressures extending up to 100~GPa, we always observe
symmetry-lowering distortions, associated to oxygen-octahedron
tiltings and the stereochemical activity of Bi. This prediction is
consistent with the experimental results of Gravriliuk {\sl et
al}.~\cite{gavriliuk-old} and Haumont {\sl et
al}.~\cite{haumont-submitted}, who observe a non-cubic phase at high
pressures and room temperature.

Finally, let us comment on the status of our {\sl quantitative}
results. As mentioned above, we repeated all our calculations for
different values of the $U$ parameter that defines the LDA+U
functional. For the $U$ values typically used in studies of
BFO~\cite{neaton05,ederer05,kornev07} and other oxides with
Fe$^{3+}$~\cite{rollmann04} (i.e., in the 3--4~eV range), the obtained
qualitative results are identical. At the quantitative level, the main
difference is a positive shift of the transition pressures as $U$
increases. For example, for $U$=4~eV the HS--LS transition occurs at
42~GPa, as compared with 36~GPa for $U$=3~eV. The differences are
greater for the LS--NS transition, which occurs at about 130~GPa for
$U$=4~eV. This is not surprising, as a bigger value of $U$ will tend
to favor more insulating solutions. Hence, we think we can take our
quantitative results as quite approximate in what regards the HS--LS
transition and the properties of the HS and LS phases. The LS--NS
transition pressure is, obviously, not well determined.

In summary, we have revealed and explained the occurrence of a
manifold (structural/electronic/magnetic) phase transition in
BiFeO$_3$ at pressures of about 40~GPa. We hope our results will
contribute to a better understanding of BiFeO$_3$'s $T$-$P$ phase
diagram and the complex interactions at work in this material.

We acknowledge fruitful discussions with G.~Catalan, J.~Kreisel, and
J.F.~Scott. This is work funded by MaCoMuFi (STREP\_FP6-03321). It was
also supported by CSIC (PIE-200760I015) and the Spanish
(FIS2006-12117-C04-01, CSD2007-00041) and Catalan (SGR2005-683)
Governments. We used the CESGA computing center.


\begin{thebibliography}{99}
\bibitem{fn-tech} We used the PAW scheme~\protect\cite{vasp}, solving for the
following electrons: Fe's 3$d$, and 4$s$, Bi's 6$s$, and 6$p$, and O's
2$s$ and 2$p$; 400~eV plane-wave cut-off; $\Gamma$-centered
7$\times$7$\times$7 k-point grid for Brillouin Zone
integrals. Calculation conditions checked to be converged. For BFO at
0~GPa we reproduce previous first-principles
results~\protect\cite{ravindran06,neaton05}.
\bibitem{fn-1200} The 1200~K transition to the cubic phase does {\sl
not} proceed from the $R3c$ phase stable at low temperatures, but from
an intermediate phase~\protect\cite{palai08}. Yet, those 1200~K {\sl
measure} the strength of the symmetry-lowering distortions in BFO, and
it seems reasonable to correlate this quantity with our computed
enthalpy difference.
\end{thebibliography}
\end{document}